\documentclass[useAMS,usegraphicx,usenatbib,float]{mn2e}
\usepackage{color}
\usepackage{float}
\usepackage{graphicx}
\usepackage{natbib}
\usepackage{chngpage}

\def\simgt{\lower.5ex\hbox{\gtsima}}

\title{SZ effect or Not? - Detecting most galaxy clusters' main foreground effect}

\author{{Weike Xiao$^{1}$\thanks{E-mail:wkxiao@hit.edu.cn}, Chen Chen$^{2}$, Bin Zhang$^{2}$, Yongfeng Wu$^{3}$\thanks{E-mail:yongfeng.wu@maine.edu}, Mi Dai$^{4}$}\\
$^{1}$Department of Astronautics Engineering, Harbin Institute of Technology, Heilongjiang Province, China, 150001\\
$^{2}$Department of Physics, Harbin Institute of Technology, Heilongjiang Province, China, 150001\\
$^{3}$American Physical Society, Maryland, USA \\
$^{4}$Homer L. Dodge Department of Physics and Astronomy, University of Oklahoma, Norman, OK 73019 USA}

\begin{document}



\maketitle

\label{firstpage}

\begin{abstract}
Galaxy clusters are the most massive objects in the Universe and comprise a high-temperature
intracluster medium of about $10^7$ K, believed to offer a main foreground effect for cosmic
microwave background (CMB) data in the form of the thermal Sunyaev¨CZel¡¯dovich (SZ)
effect. This assumption has been confirmed by SZ signal detection in hundreds of clusters
but, in comparison with the huge numbers of clusters within optically selected samples from
Sloan Digital Sky Survey (SDSS) data, this only accounts for a fewper cent of clusters. Here
we introduce a model-independent new method to confirm the assumption that most galaxy
clusters can offer the thermal SZ signal as their main foreground effect.For the $Wilkinson
Microwave Anisotropy Probe$ (WMAP) seven-year data (and a given galaxy cluster sample),
we introduced a parameter $d_1$ as the nearest-neighbour cluster angular distance of each pixel,
then we classified data pixels as ¡®to be¡¯ ($d_1\rightarrow 0$ case) or ¡®not to be¡¯ ($d_1$ large enough) affected
by the sample clusters. By comparing the statistical results of these two kinds of pixels, we
can see how the sample clusters affect the CMB data directly. We find that the $Planck$ Early
Sunyaev¨CZel¡¯dovich (ESZ) sample and X-ray samples (¡«$10^2$ clusters) can lead to obvious
temperature depression in the $WMAP$ seven-year data, which confirms the SZ effect prediction.
However, each optically selected sample ($>10^4$ clusters) shows an opposite result: the mean
temperature rises to about $10\mu$ K. This unexpected qualitative scenario implies that the main
foreground effect of most clusters is $not$ always the expected SZ effect. This may be the reason
why the SZ signal detection result is lower than expected from the model.

\end{abstract}


\section{Introduction}

As the most massive self-gravitating systems in the cosmos, galaxy
clusters can make significant contributions to the measurements
of precision cosmology and, during their formation and evolution,
various effects can be analysed statistically. One such major effect
is the Sunyaev¨CZel¡¯dovich (SZ) effect (\cite{SZ}): cosmic microwave background (CMB) photons can undergo
inverse Compton scattering off high-energy electrons in the intracluster
medium (ICM) when passing through clusters. The thermal
SZ effect is considered to be a most remarkable effect, as it can
increase photon energy statistically and noticeably distort the CMB
spectrum. The thermal SZ effect decreases intensity at low frequencies
(like the Q, V and W bands of WMAP) and increases intensity
at high frequencies.

This predictable distortion of CMB data is recognized as a marked
signal of clusters. After the SZ signal of residential clusters was
confirmed, blind sky surveys using the SZ effect continued over the
last decade, including those of the South Pole Telescope (SPT:\cite{a3}), Atacama Cosmology Telescope (ACT:
\cite{a4}), Planck (\cite{a5}) and others. A number of high-redshift clusters are expected to be found
via blind SZ sky surveys using features of the SZ signal and its insensitive
properties with red shift (\cite{a2}).
This provides a solid foundation for the measurment requirements of precision cosmology.

Here we note that a basic assumption is behind the expectations of
these ongoing projects: the main foreground effect of most galaxy
clusters on CMB data should be the expected thermal SZ effect.
Although SZ signals of hundreds of clusters have already been
detected and confirmed, we cannot assume that such signals exist
for all clusters. In comparison with the huge cluster numbers in
optically selected samples from Sloan Digital Sky Survey (SDSS) data, the basic assumption has only been validated in a few percent
of clusters (and it is hard to provide their selection function: see
 \cite{a11}). Considering that unknowns can have
unpredictable or unforeseen impacts on understanding or applying
the results of SZ effect galaxy cluster surveys, we therefore need
a means of direct detection (but not using 1.4-GHz data to analyse
the 150-GHz case) of most (large percent) clusters to confirm this assumption.

At the same time, cosmological analysis methods such as the 'luster number count'(\cite{a6,a7}) expect a complete catalogue of setting conditions. It is
important to ensure that SZ-effect blind surveys of galaxy clusters do
not miss cluster samples.However, some studies have already found
the observed SZ-effect signal to be (especially for optically selected
clusters) not strong enough (\cite{a9,a10,a13c,a12,a13,a13b}), although this is still in
debate (\cite{a19,a23,a24}). The existence and negligibility of SZ-effect signals
is becoming a noteworthy debate. We note that some traditional
analysis methods are model-dependent and that the free parameters
can lead to uncertainty in the debate. It is therefore necessary to
introduce a new model-independent method that will ensure more
reliable conclusions can be drawn.

\section{Methods}

To study foreground effects of galaxy clusters, one can consider the
viewpoint of CMB data pixels, simply taking each pixel as a probe.
For one galaxy cluster in an ideal isotropy CMB, a simple method
is used to compare the probe data (temperature data of this pixel)
of angular regions affected and unaffected by the cluster. For real
CMB data, the fluctuation temperature of each pixel can be taken
as another Gaussian distribution error of the detector. Considering
the different properties of noise signals and the SZ signal, one can
use statistical methods to compare the mean probe data of angular
regions considered 'to be' or 'not to be' affected by the sample
clusters. The noise signal will have similar effects on these two
kinds of pixels, but the thermal SZ signal will only depress the
temperature of 'to be' affected pixels.

The preconditions here are twofold: first that we are able to
differentiate these two angular regions and secondly that each region
includes enough data pixels to minimize the statistical error.

\begin{figure}
    \includegraphics[height=0.5\textwidth, angle=0]{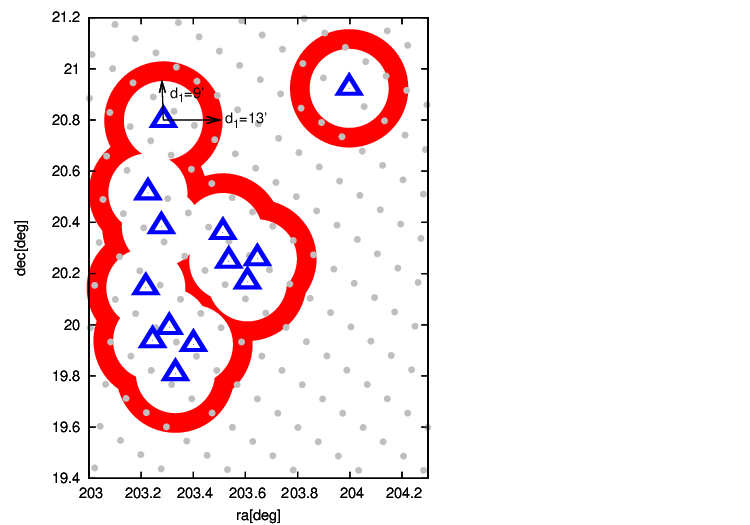}
  \caption{\sffamily\textbf{Figure 1.}
    Pixel region round a group of clusters within a bin value of $
9 \arcmin < d_1 < 13 \arcmin$. Each triangle corresponds to the position of
one cluster in the GMBCG sample and each grey disc denotes the central
position of one pixel in the WMAP data. For the sparse cluster sample (the
top right one), we can see that our method will simply retrogress to the
stacking method.
   \label{fig:fig1}}
\end{figure}

We find that a continuous parameter $d_1$ is competent for such
taxology. For each CMB data pixel, $d_1$ is defined as the angular
distance of this pixel to its nearest-neighbour galaxy cluster (of
the cluster sample used). Comparison with the traditional 'stacking
method' can help us to understand the parameter $d_1$. The stacking
method selects each cluster as the origin and bins pixels by circular
rings around it (with different angular distances), using the stacking
annulus of all clusters to perform analysis. Considering that galaxy
clusters tend to swarm together, the temperature signal of pixels
within an annulus around one cluster can be seriously affected by
neighbouring clusters. In Fig.\ref{fig:fig1}, we show the pixel region around a
group of clusters within a bin value of $d_1$. Our method can be comprehended
as changing the circle annulus around each cluster with
curved loops around the collection of sample clusters and redefining
the angular parameter as $d_1$. This means that the angular distance
to the whole cluster sample is used, rather than those for each cluster.

As clusters mainly affect their local angular region, pixels with
$d_1$ parameter large enough can properly represent the 'not to be'
affected angular regions, unlike the stacking method. One other
technical merit here is the impartial use of pixel data, as the stacking method applies pixels affected by neighbour clusters more times.
When the cluster sample is sparse in the sky, our method simply
reverts to the stacking method.

\begin{figure}
    \includegraphics[height=0.5\textwidth, angle=270]{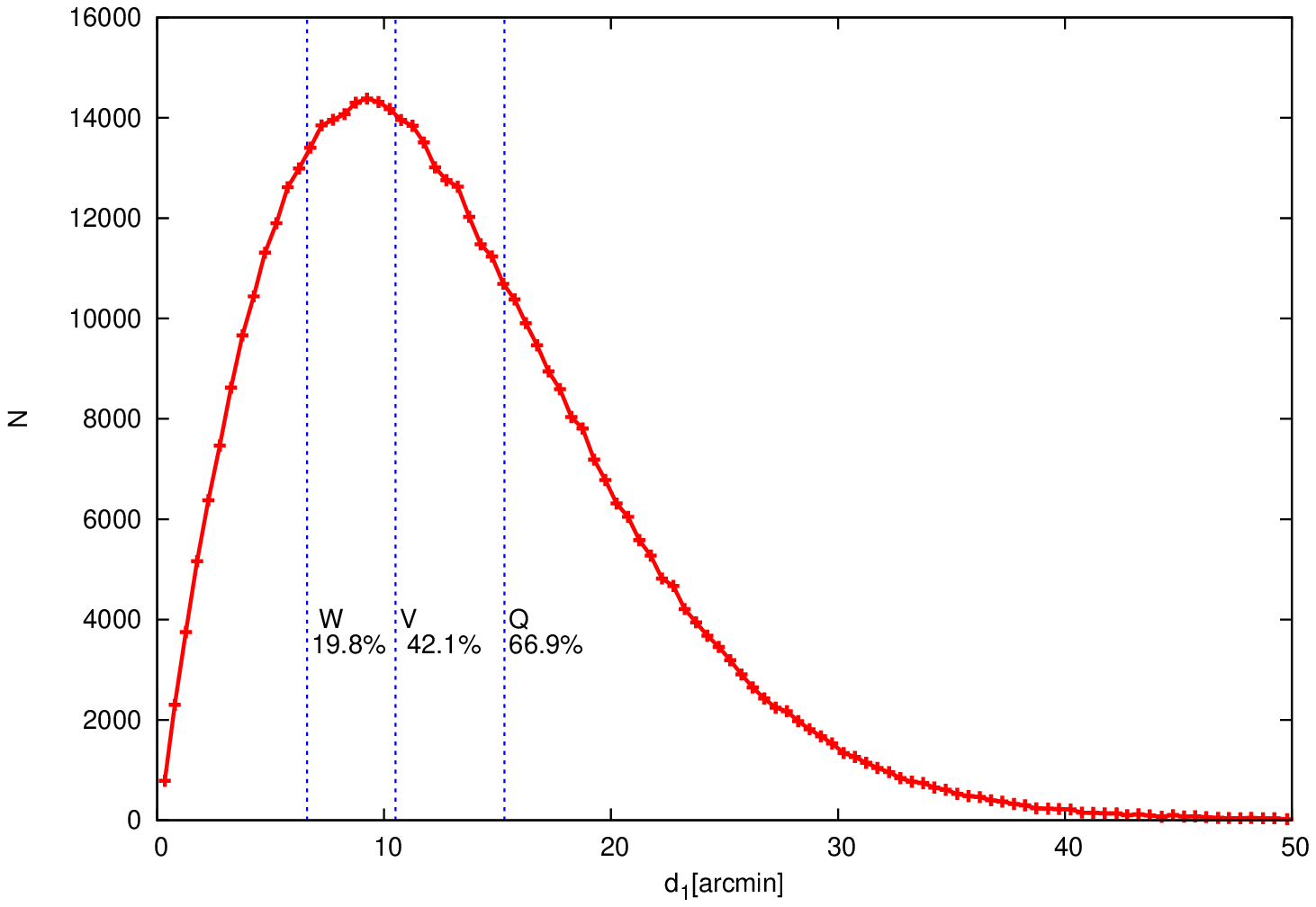}
  \caption{\sffamily\textbf{Figure 2.}  The distribution function of pixel $d_1$ parameter. The pixels correspond
to the CMB data within the main angular region of the GMBCG
sample. We select the 50 580 clusters of the sample in the main survey
area of SDSS DR-7 and take 503 740 $WMAP$ data pixels within the area
as our statistical CMB angular region. In order to avoid some unwanted
edge effects, we have already dropped pixels close to the edge within
70 arcmin. The dotted lines (blue in the online article) show the beam size
of $WMAP$ data in the Q, V and W band, respectively, and the percentages are
the proportions of pixels with $d_1$ values within these half beam sizes.
   \label{fig:fig2}}
\end{figure}

Here we illustrate the results with the optically selected Gaussian
Mixture Brightest Cluster Galaxy (GMBCG:\cite{a14}) galaxy
cluster sample. In Fig.\ref{fig:fig2}, we show the distribution function of the
pixel $d_1$ parameter. This figure shows that the effective statistical
region is within $ 0<d_1<40'$.By comparing the possible angular size
within which one cluster might affect pixel data (mainly the beam
size), one can see that both cluster-affected and cluster-unaffected
regions contain enough pixels for effective statistical analysis. In
contrast, due to the angular resolution of $WMAP$ data, this is difficult
to do with SDSS galaxy samples.

We can analyse the  $<T>$--$<d_1>$ curve (hereafter TD curve) by taking
each pixel as a probe and comparing the mean temperature $<T>$ of
pixels binned with different $d_1$ values. The merits of this method
will be discussed in detail in another article. Here, we emphasize
two points relating to the physics.

First, the two sides of the TD curve represent different cases
of pixels being affected or unaffected (by the cluster sample used).
The main foreground effects of galaxy clusters that we are interested
in (such as the SZ effect and radio emission) have the property of
'angular localization', which means that they only affect the angular
region they appear in (within several arcmin for most clusters). Since
the beam angle of $WMAP$ data in Q, V and W bands and Internal
Linear Combination (ILC) data ranges from 13 arcmin to about
30 arcmin, it is safe to say that pixels of  $d_1 \rightarrow 40'$ represent
angular regions 'surely unaffected' and that $d_1\sim 0'$ pixels
represent 'affected regions', judging by the cluster samples. By
comparing the mean temperatures of these two cases, we can see
how galaxy clusters affect CMB data directly.

Secondly, some background or foreground unrelated objects outside
the cluster sample might also affect the CMB data, but statistically
they will change both sides of the TD curve in the same
way. This can be tested by simulation. Thus, if a reliable difference
between the two sides is confirmed, it should be an effect caused by
the cluster sample itself.


\begin{figure}
  \begin{center}
    \includegraphics[height=0.45\textwidth, angle=270]{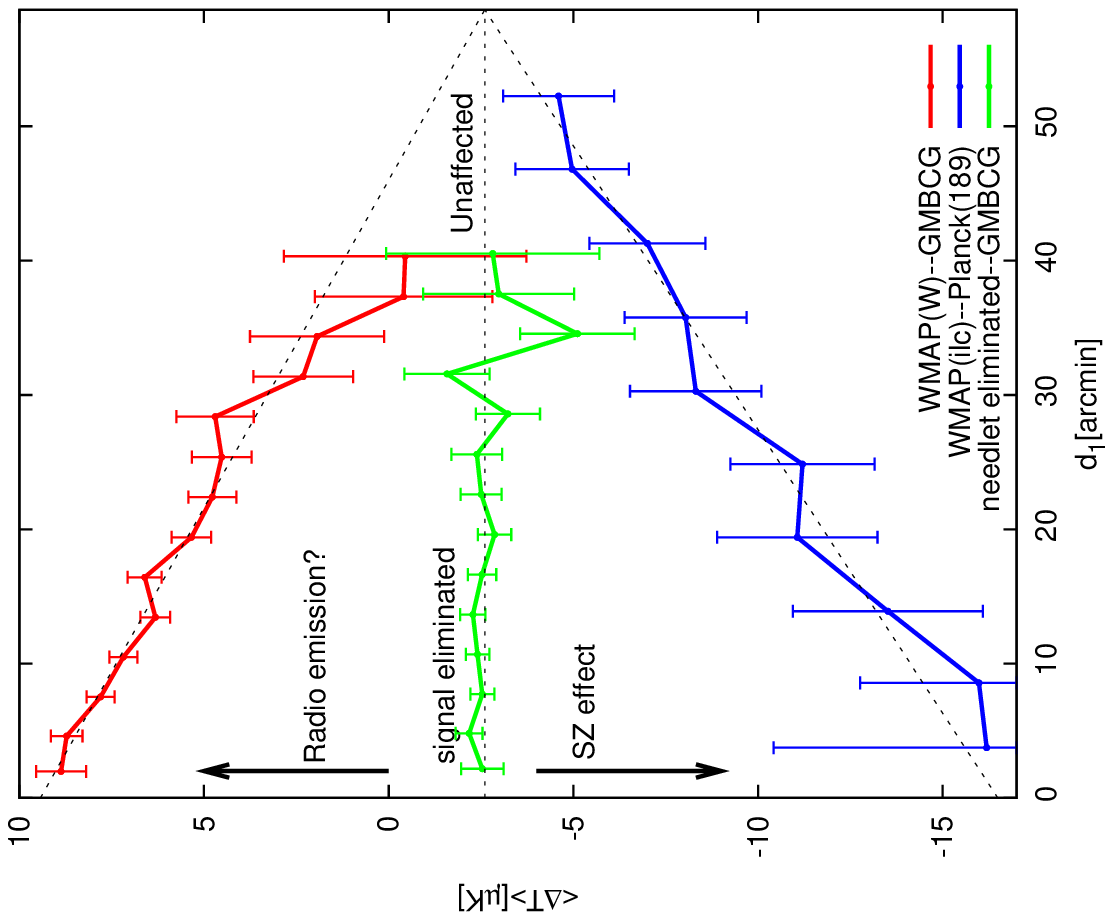}
  \end{center}
  \caption{\sffamily\textbf{Figure 3. } Comparison of the mean temperature of pixels with different $d_1$ parameters. The figure shows typical TD curves of different CMB data.galaxy cluster samples. The temperature distribution function of pixels within each $d_1$ bin (3') can be well described with a Gauss function and we calculate the error bar of$<T>$ with $\sigma=\sqrt{\frac{\sum(T_i-<T>)^{2}}{N(N-1)}}$ (N is the pixel number within the $d_1$ bin). The curves can be described with an empirical line function: $<T>=A * <d_1> + T_1$ (fitting with weight $\sqrt{N}$). The traditional stack method is somehow our special case when the cluster sample is sparse and the neighbour-cluster effect can be neglected, as is the case for the Planck-189 sample. For the Planck-189 sample, we select $WMAP$ data pixels round each cluster within 55' as our statistical region (the bin width is 5.5'). Since only the slope $A$ is really interesting, the $WMAP$(ILC).$Planck$(189) cluster TD curve is translated as $<T>-13uK$ here, to show more clear contrast with the same $<T>$ in the unaffected region.
   \label{fig:fig3}}
\end{figure}

Nowwe can see the qualitative scenario as regards how the cluster sample affects the TD curve. When the thermal SZ effect is the main foreground effect, the mean temperature of $WMAP$ data in Q, V and W bands should decline noticeably for cluster regions ($d_1\rightarrow 0'$) and the large  $d_1$ side should remain unaffected; then, the TD curve should rise from $d_1=0'$ to $d_1=40'$. In contrast, if radio emission were the main foreground effect then the low $d_1$ side would be driven up and the TD curve would decline.

\section{Results}

In Fig.\ref{fig:fig3}, we illustrate the TD curve of the optically selected GMBCG sample (\cite{a14}) and 189 Planck SZ-effect selected clusters (Planck-189:\cite{a11}). For the Planck-189 sample, the TD curve line is obviously rising, which means  $<T>$ is much lower when $d_1\rightarrow 0'$ than in the large $d_1$ region, thus indeed confirming the SZ-effect prediction. However, when the foreground is changed to the 50 580 clusters of the GMBCG sample, we obtain an unexpected opposite result: the TD curve becomes a visible downward curve,which means that the mean temperature behind clusters has an increment up to about $10\mu$K. This result is similar for $WMAP$ data in Q, V and W bands and also the ILC data, with small error margins (while the TD curve becomes a nearly level line when we use the CMB data eliminated by another
team (\cite{a20}) using the needlet method.)

\begin{figure}
  \begin{center}
    \includegraphics[height=0.35\textwidth, angle=0]{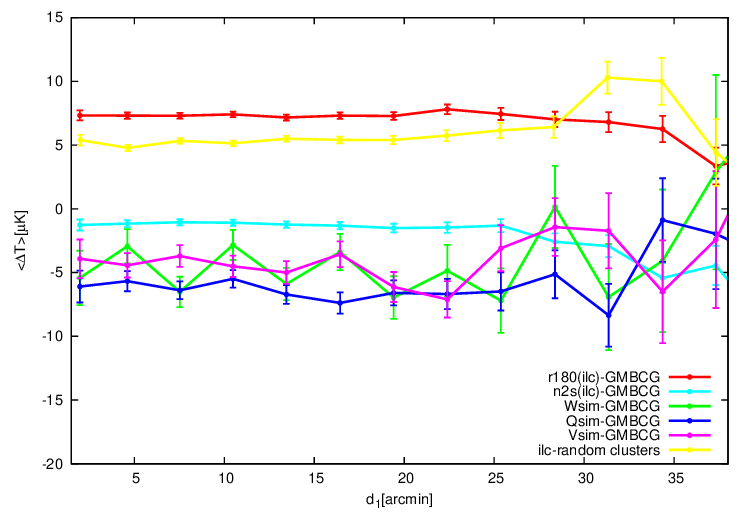}
  \end{center}
\caption{\sffamily\textbf{Figure 4.}  The TD curves of testing data of CMB and clusters. This figure includes (1) a randomized cluster sample (of GMBCG); (2) simulated CMB data in Q, V and W bands; (3) a reversal of CMB ILC data r180 ($ ra\rightarrow ra+180^\circ$), n2s ($dec\rightarrow -dec$). For all the CMB data, we select the
503 740 pixels within GMBCG's main region. (Since we are not using all-sky pixels, we note that their mean temperature is not zero.)
    \label{fig.fig4}}
\end{figure}

As a post hoc examination, we also calculated the TD curves using the simulated CMB data of $WMAP$ and also the simulated random distribution cluster sample. For these, no causal relationship data were found; Fig.\ref{fig.fig4} shows that their TD curves are common, level curves as expected.

Fig. \ref{fig:fig3} also shows us that the TD curves of different samples are typically approximately straight lines, so we can describe them with an empirical line function:

$$<T>=A * <d_1> + T_1$$

If we are not using one all-sky cluster sample, the parameter $T_1$ can be influenced by the cluster sample region and also large-scale CMB fluctuation. The $A$ slope is a valuable parameter. For the samples we used, the value $\Delta T_A\equiv -A*40'$ can show us roughly the difference of $<T>$ between affected and unaffected regions, which underlies how these cluster samples affect CMB data. $\Delta T_A<0$ corresponds to the results of the thermal SZ effect and $\Delta T_A>0$ relates to an opposite effect like radio emissions.

In Table \ref{tab:results} we show the value of $\Delta T_A$ when setting different cluster samples as the foreground of the $WMAP$ seven-year data (\cite{a8}: W band and ILC data). Here we can see the $\Delta T_A $ values fall explicitly into two situations: for SZ-effect-selected and X-ray-selected cluster samples (ACT: \cite{a21}; Planck-189:\cite{a11}; XMM Cluster Survey (XCS-DR1):\cite{a22}; Meta-Catalogue of X-ray detected Clusters of galaxies (MCXC): \cite{a16})) $\Delta T_A $ is obviously negative, confirming the SZ-effect image; yet for each optically selected cluster sample (GMBCG: \cite{a14}; Wen: \cite{a18}; maxBCG: \cite{a17}) the $\Delta T_A $ value is significantly positive.

\begin{table}
\center

\begin{tabular}{r@{--}lrrr}
\hline
CMB&cluster  & $N_{cluster}$  & $\Delta T_A\equiv-A*40'$       & $\Delta$A/A\\
\hline
WMAP(W)&GMBCG     & 50580 & 8.2 $\mu$K& 6.9 \% \\
WMAP(V)&GMBCG     & 50580 & 6.5 $\mu$K& 8.9 \% \\
WMAP(Q)&GMBCG     & 50580 & 7.3 $\mu$K& 4.5 \% \\
WMAP(ilc)& GMBCG  & 50580 & 6.9 $\mu$K& 5.0 \% \\
WMAP(W)&Wen       & 83279 & 6.3 $\mu$K& 20.3 \% \\
WMAP(W)&maxBCG    & 13823 & 5.3 $\mu$K& 21.7 \% \\

\hline

WMAP(W)& ACT          & 23     & -19.4 $\mu$K& 25.1 \% \\
WMAP(ilc)& ACT        & 23     & -11.5 $\mu$K& 17.3 \% \\
WMAP(W)& Planck(189)  & 189    & -23.5 $\mu$K& 25.0 \% \\
WMAP(ilc)& Planck(189)& 189    & -9.5 $\mu$K& 7.0 \% \\
WMAP(W)& xcs3         & 503    & -7.9 $\mu$K& 32.2 \% \\
WMAP(ilc)& xcs3       & 503    & -6.8 $\mu$K& 15.3 \% \\
WMAP(W)& MCXC         & 1743   & -7.7 $\mu$K& 23.6 \% \\
\hline
\end{tabular}

\caption{\label{tab:results} Fitting results of different samples. }
\label{table}
\end{table}


In summary, with the statistical method of $d_1$ parameters we can sum up the experiential foreground effect of galaxy clusters in these qualitative points.

(i) In angular regions of galaxy clusters, we can see the main foreground effect is NOT the thermal SZ effect, but rather there exists an opposite contamination foreground effect somewhat like radio emission. Such an opposite signal in most clusters is high enough that it can cover the SZ-effect signal of the cluster and act as the main foreground effect. It should not be neglected when performing CMB signal analysis (whether the cluster samples are surely clusters or not).

(ii) With regard to distance, this contamination should come from the cluster itself, because if it affects the line of sight before or after the cluster, such as is the case for a star burst galaxy at high redshift background, then statistically there should also be the same effect in non-cluster regions and a temperature change should not result.

(iii) In spectra, such 'emission components' show similar effects in Q, V and W bands, a little lower at the V frequency.

\section{Discussion and conclusions}

Before performing a quantitative study about $\Delta T_A $ values in more detail, we focus on the explicit $\Delta T_A >0$ property of each optically selected cluster sample. Its value is about zero when we use a randomly distributed foreground sample in Fig.\ref{fig.fig4}, so it is interesting to understand why $\Delta T_A $ changes in Fig. \ref{fig:fig3}.

\begin{figure}
  \begin{center}
    \includegraphics[height=0.5 \textwidth, angle=270]{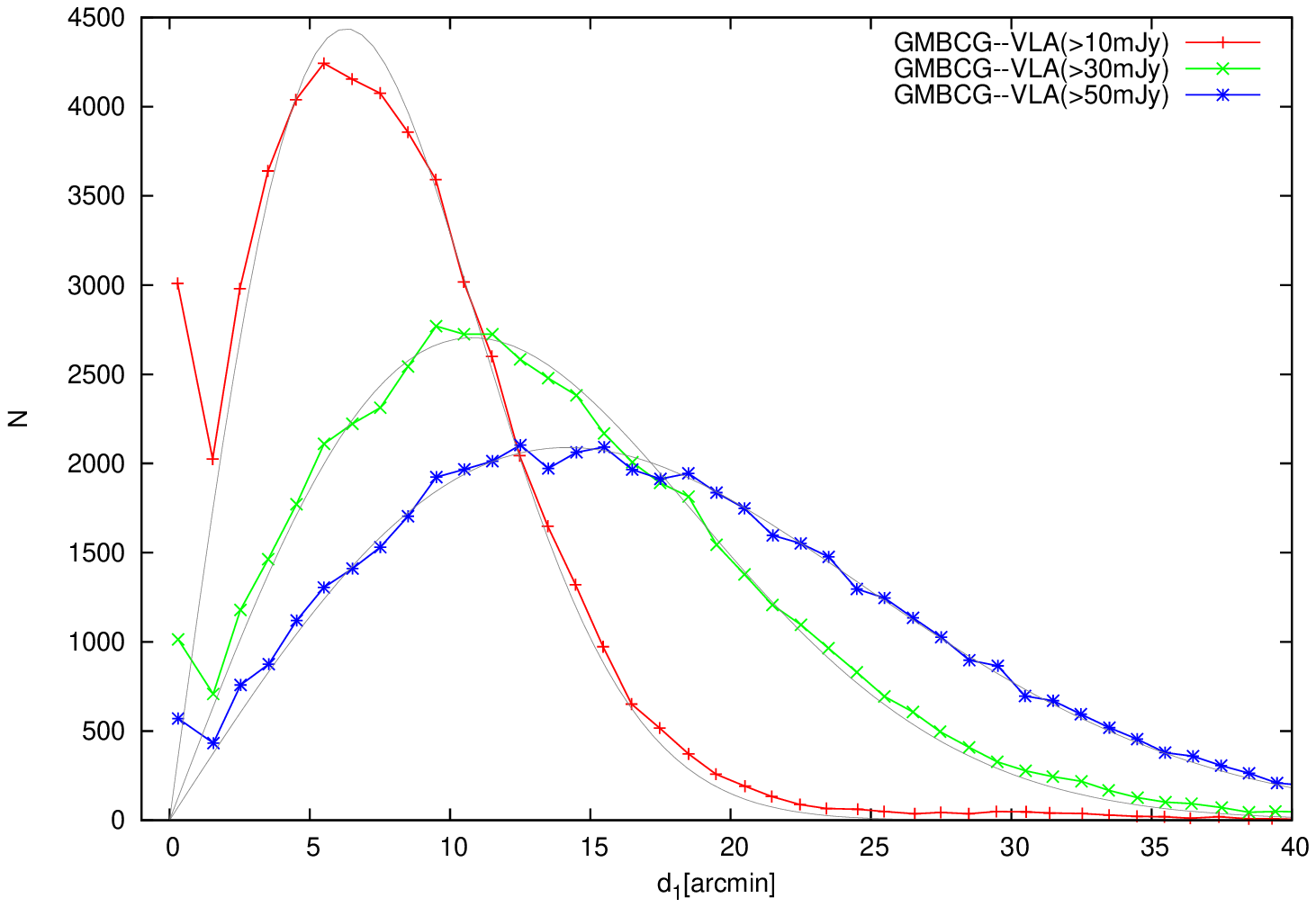}
  \end{center}
\caption{\sffamily\textbf{Figure 5. }  The $d_1$ distribution curves of GMBCG sample with respect to VLA samples. Here $d_1$ is defined as the angular distance of each GMBCG cluster's nearest-neighbourhood VLA FIRST Survey (1.4 GHz) sample.We choose flux threshold cuts of 50,30 and 10 mJy at 1.4 GHz to yield bright radio sources at high frequency. We also fit the $d_1 > 1'$ bins with a model of random distribution in VLA sample number density.
    \label{fig.fig5}}
\end{figure}

One might imagine that a few bright radio point sources within these samples make the mean temperature positive. Traditional cross-correlation analyses (\cite{a13b}) have found that only a few percent of clusters might be affected by a nearby radio source. Here we show similar results with $d_1$ parameter analysis between the GMBCG sample and the radio sources of Very Large Array (VLA) Faint Images of the Radio Sky at Twenty centimetres (FIRST) catalogues \footnote{http://sundog.stsci.edu/first/catalogs.html} (\cite{FIRST}). In the $d_1 \rightarrow 0$ case, Fig. \ref{fig.fig5} shows that some radio sources are surely at the cluster region, but only a few percent in total. The results for the $d_1 > 1'$ part can be well-fitted with a random-distribution model of VLA source number density. Remember that such randomly distributed sources affect both sides of the TD curve in the same way, so radio-loud sources will not cause a $\Delta T_A$ change.

Here we suggest that this is the result of galaxy radio emissions. The TD curve compares angular regions that are 'galaxy-rich' (cluster region) and 'hold few galaxies' (no-cluster region), so it can also represent such emissions. Each galaxy is also a diffuse foreground object affecting the observation results with similar spectra like the Galaxy (see fig. 22 in the WMAP nine-year data: \cite{wmap9}). Such models suggest an antenna temperature increase of round about $10\mu$K in Q, V and W bands, while the V-band result is smaller, confirming the results in our table. This model also suggests that thermal dust emission will be most serious in the 150-GHz case and we will see similar and slightly higher $\Delta T_A$ for future Planck and ACT data.


However, the discussion above did not consider the cluster SZ effect. If each cluster can offer a typical SZ signal of more than $100\mu$K antenna temperature decrease in the W band, the above effect can be neglected, yet the TD curves of optically selected clusters in Fig. \ref{fig:fig3} have shown the opposite result and negated this
point. A possible conclusion is therefore that only a few clusters can offer a SZ signal (or most of them are very weak and can be covered by the radio-emission effect of galaxies). The reason may be simple: the ICM reaches high energy just after cluster formation and can then offer both thermal SZ effect and X-ray emission; however, such an effect cannot last for a long time-scale since the ICM is losing energy in a major way. This scenario clarifies why the SZ effect signal is apparently weaker than expected (\cite{a11,a12,a13,a13b}):

As regards SZ signal blind surveys or X-ray observations, these observe clusters that are able to offer a relative signal in the sky. With this selection effect, the observed samples are all within the ICM high-energy time-scale, so the detected cluster number will be small and the statistical result will show high SZ or X-ray signal, such as is the case for the TD curve of the Planck-189 sample in Fig. \ref{fig:fig3}.

For optically selected galaxy clusters, the optical method gives a much more complete cluster sample, including a large percentage of clusters outside the ICM high-energy time-scale. In this case, the ICM of most clusters is not at high energy and its SZ signal is weak, so the main foreground effect can be covered by galaxy radio emission. We can thus see the $\Delta T_A >0$ effect in Fig. \ref{fig:fig3} and the lower $Y_{500}$ signal for optically selected samples.

In conclusion, our model-independent method shows the main foreground effect of most (but not a few hundred) galaxy clusters directly (but not using the 1.4-GHz data imaging W-band result). The results of known SZ-signal-selected clusters and X-ray-observed clusters confirm the traditional thermal SZ result. Unexpectedly, however, the thermal SZ signal of most clusters in optically selected samples is contaminated (even covered) by something like radio emission. This may be the reason why the SZ signal detection result is lower than model expectations.

\section*{Acknowledgments}
The authors sincerely thank Prof. Liu X.W.'s support in KIAA-PKU.
The project is supported by Key Laboratory Opening Funding of Technology of Micro-Spacecraft
(HIT.KLOF.2009098) and also by the development program for outstanding young teachers in HIT (BAQQ 92324501).



\bsp

\label{lastpage}

\end{document}